# Radiation of a neutral polarizable particle moving uniformly through a thermal radiation field

G.V. Dedkov and A.A. Kyasov

Nanoscale Physics Group, Kabardino-Balkarian State University, Nalchik, Russia

**Abstract** –We discuss the properties of thermal electromagnetic radiation produced by a neutral polarizable nanoparticle moving with an arbitrary relativistic velocity in a heated vacuum background with a fixed temperature. We show that the particle in its own rest frame acquires the radiation temperature of vacuum, multiplied by a velocity-dependent factor, and then emits thermal photons predominantly in the forward direction. The intensity of radiation proves to be much higher than for the particle at rest. For metal particles with high energy, the ratio of emitted and absorbed radiation power is proportional to the Lorentz-factor squared.

## 1. Introduction

Since the pioneering works by Lebedev on measuring the pressure of light (1899) and Planck on the quantum nature of electromagnetic radiation and thermal radiation emitted by heated bodies (1900), the problem of the interaction of electromagnetic field with matter has been in the focus of many researchers in the past century and up to the present time. In particular, the questions relating to the confusing transformations of temperature and spectral distributions of thermal radiation of large-size bodies in different inertial reference systems are intensively discussed so far (see [1--4] and references).

In this work, we attack the problem of thermal radiation emitted by a neutral moving particle of a small size, relaxing the assumption of thermal and dynamical equilibrium between the particle and vacuum environment. Since the conditions of geometrical optics are not fulfilled for the particles with a size lesser than the Wien wavelength, the formalism which is used to determine the drag caused by the black body radiation [5] is not applicable in the problem of thermal radiation of such particles. We consider two inertial reference frames, one of which is co-moving with the particle (own reference frame) and the other one is related with the reference frame of vacuum (thermalized photonic gas). We also assume that the particle and vacuum are characterized by different local temperatures which are well defined in the corresponding reference frames. As a result, we obtain a full set of equations describing the dynamics of relativistic particle, its thermal state in the co-moving reference frame and the intensity of radiation in the reference frame of vacuum. The particle emission turns out to be considerably higher than absorption depending on the Lorentz-factor. Thermal radiation is concentrated within a narrow cone within the direction of the particle velocity, while its intensity is governed by the temperature of the background and dielectric properties of the particle. To demonstrate these results, we have carried out numerical calculations in the case of metallic (gold) particles with a size of 5 and 50 $nm$.

## 2. Theory

Consider a small particle of radius $R$ uniformly moving with velocity $V$ through a thermalized photonic gas with temperature $T_2$ (fig. 1). Let the surface $\sigma$ encircles the particle at a large enough distance so that the fluctuation electromagnetic field on $\sigma$ represents the wave field. The reference frames $\Sigma$ and $\Sigma'$ correspond to the reference frames of the photonic gas (background radiation) and particle, respectively. Initially, the particle has the temperature $T_1$ in its own reference frame. We also assume the validity of conditions $R << \min(2\pi\hbar c/k_B T_1, 2\pi\hbar c/k_B T_2)$. In this case, when emitting thermal photons, the particle can be considered as a point-like dipole with fluctuating dipole and magnetic moments $\mathbf{d}(t), \mathbf{m}(t)$, and its material properties are described by the frequency-dependent dielectric and (or) magnetic polarizabilities $\alpha_e(\omega)$, $\alpha_m(\omega)$.

According to the conventional form of the energy conservation law in the system (the system of vacuum in this case) within the volume $\Omega$ restricted by the external closed surface $\sigma$ we may write [6]

$$-\frac{dW}{dt} = \oint_\sigma \mathbf{S} \cdot d\vec{\sigma} + \int_\Omega \langle \mathbf{j} \cdot \mathbf{E} \rangle d^3 r , \qquad (1)$$

where

$$W = \int_\Omega \frac{\langle \mathbf{E}^2 \rangle + \langle \mathbf{H}^2 \rangle}{8\pi} d^3 r , \qquad (2)$$

and

$$\mathbf{S} = \frac{c}{4\pi} \langle \mathbf{E} \times \mathbf{H} \rangle . \qquad (3)$$

denote the energy of fluctuating electromagnetic field in the volume $\Omega$ and the Poynting vector of this field. The second term in (1) represents the Joule energy dissipation integral, and the angular brackets in (1)—(3) denote total quantum and statistical averaging. Within the quasistationary approximation used, $dW/dt = 0$, and from (1) we obtain the general expression for the intensity of radiation

$$I = \oint_\sigma \mathbf{S} \cdot d\vec{\sigma} = -\int_\Omega \langle \mathbf{j} \cdot \mathbf{E} \rangle d^3 r \equiv I_1 - I_2, \tag{4}$$

where $I_1 = I_1(T_1)$ determines the intensity of thermal radiation emitted by the particle in vacuum, and $I_2 = I_2(T_2)$ is the absorbed intensity of the background radiation.

Our next step is to represent the right-hand side of Eq. (4) in a more convenient form. In the case of stationary electromagnetic fluctuations, using relativistic transformations for the current density, electric field and volume in systems $\Sigma$ and $\Sigma'$, we obtain ($\beta = V/c$) [7,8]

$$\int_{\Omega'} \langle \mathbf{j'} \cdot \mathbf{E'} \rangle d^3 r' = \frac{1}{1-\beta^2} \left( \int_\Omega \langle \mathbf{j} \cdot \mathbf{E} \rangle d^3 r - F_x \cdot V \right) \equiv \frac{1}{1-\beta^2} dQ/dt, \tag{5}$$

$$dQ/dt = \langle \mathbf{d} \cdot \mathbf{E} + \dot{\mathbf{m}} \cdot \mathbf{H} \rangle, \tag{6}$$

$$\frac{dQ'}{dt'} \equiv \int_{\Omega'} \langle \mathbf{j'} \cdot \mathbf{E'} \rangle d^3 r'. \tag{7}$$

$$\mathbf{F} = \int \langle \rho \mathbf{E} \rangle d^3 r + \frac{1}{c} \int \langle \mathbf{j} \times \mathbf{H} \rangle d^3 r = \langle \nabla (\mathbf{d} \cdot \mathbf{E} + \mathbf{m} \cdot \mathbf{H}) \rangle. \tag{8}$$

In Eq. (5), $F_x$ is the projection of $\mathbf{F}$ onto the velocity direction (only this component of force differs from zero in this case). The points above the dipole moments in (6) denote the time derivative. Obviously, $dQ'/dt'$ determines the heating rate of the particle in the co-moving frame of reference.

Using (4), (5) yields

$$I = I_1 - I_2 = -\left( \frac{dQ}{dt} + F_x V \right), \tag{9}$$

where $dQ/dt$ and $F_x$ are given by Eqs. (6) and (8). The relationship between $dQ'/dt'$ and $dQ/dt$ through (5) and (7) formally corresponds to the Planck formulation of relativistic thermodynamics assuming that $Q$ is associated with the amount of heat in the reference frame $\Sigma$. However, this is only an ad hoc assertion. At this point, really important for us is the quantity $Q'$ which is a priori related to the amount of heat in $\Sigma'$. This allows one to calculate the temperature $T_1$ in $\Sigma'$ during the particle motion. Thus, in principle, our further results may shed some light to the problem of relativistic temperature transformation.

Equations (5)—(9) are the basis in our further calculations of the intensity of thermal radiation. In configuration shown in Fig. 1, the general expressions for $\dot{Q}$ and $F_x$ were obtained in our previous works [9, 10] within the framework of fluctuation electrodynamics. Following our method of calculation, the quantities on the right-hand sides of (6) and (8) are written as the products of spontaneous and induced quantities (sp, ind), namely

$$dQ/dt = \left\langle \dot{\mathbf{d}}^{\text{sp}}\mathbf{E}^{\text{ind}} + \dot{\mathbf{d}}^{\text{ind}}\mathbf{E}^{\text{sp}} + \dot{\mathbf{m}}^{\text{sp}}\mathbf{H}^{\text{ind}} + \dot{\mathbf{m}}^{\text{ind}}\mathbf{H}^{\text{sp}} \right\rangle,$$

$$F_x = \left\langle \partial_x \left( \mathbf{d}^{\text{sp}}\mathbf{E}^{\text{ind}} + \mathbf{d}^{\text{ind}}\mathbf{E}^{\text{sp}} + \mathbf{m}^{\text{sp}}\mathbf{H}^{\text{ind}} + \mathbf{m}^{\text{ind}}\mathbf{H}^{\text{sp}} \right) \right\rangle.$$

All the quantities in these equations have to be Fourier-transformed over the time and space variables $t, x, y, z$. The points above the dipole moments indicate the time differentiation. The induced dipole moments $\mathbf{d}^{\text{ind}}, \mathbf{m}^{\text{ind}}$ have to be expressed through the fluctuating fields $\mathbf{E}^{\text{sp}}, \mathbf{H}^{\text{sp}}$ via linear integral relations. The induced fields $\mathbf{E}^{\text{ind}}, \mathbf{H}^{\text{ind}}$ have to be expressed through $\mathbf{d}^{\text{sp}}, \mathbf{m}^{\text{sp}}$ when solving the Maxwell equations containing the fluctuating currents induced by the dipole moments $\mathbf{d}^{\text{sp}}, \mathbf{m}^{\text{sp}}$. The arising correlators of the dipole moments and fields are calculated with the help of fluctuation-dissipation relations. Omitting the details, the resultant expressions have the form [9, 10]

$$F_x = -\frac{2\hbar\gamma}{\pi c^4} \int_0^\infty d\omega\, \omega^4 \int_{-1}^1 dx\, x(1+\beta x)^2 \alpha''(\omega\gamma(1+\beta x)) \cdot$$
$$\cdot \left[ \frac{1}{\exp(\hbar\omega/k_B T_2)-1} - \frac{1}{\exp(\hbar\omega\gamma(1+\beta x)/k_B T_1)-1} \right], \qquad (10)$$

$$\dot{Q} = \frac{2\hbar\gamma}{\pi c^3} \int_0^\infty d\omega\, \omega^4 \int_{-1}^1 dx\, (1+\beta x)^3 \alpha''(\omega\gamma(1+\beta x)) \cdot$$
$$\cdot \left[ \frac{1}{\exp(\hbar\omega/k_B T_2)-1} - \frac{1}{\exp(\hbar\omega\gamma(1+\beta x)/k_B T_1)-1} \right], \qquad (11)$$

where $\gamma = (1-\beta^2)^{-1/2}$ and $\alpha''(\omega)$ denote the imaginary part of the sum $\alpha(\omega) = \alpha_e(\omega) + \alpha_m(\omega)$. It is worth noting that a nonrelativistic limit of Eq. (10) was first obtained in [11], while the relativistic result was also derived in [12]. According to (9)—(11), one can write

$$I = -\frac{2\hbar\gamma}{\pi c^3} \int_0^\infty d\omega\, \omega^4 \int_{-1}^1 dx\, (1+\beta x)^2 \alpha''(\omega\gamma(1+\beta x)) \cdot$$
$$\cdot \left[ \frac{1}{\exp(\hbar\omega/k_B T_2)-1} - \frac{1}{\exp(\hbar\omega\gamma(1+\beta x)/k_B T_1)-1} \right], \qquad (12)$$

and

$$I_1(T_1) = \frac{2\hbar\gamma}{\pi c^3} \int_0^\infty d\omega\, \omega^4 \int_{-1}^1 dx\, (1+\beta x)^2 \alpha''(\omega\gamma(1+\beta x)) [\exp(\hbar\omega\gamma(1+\beta x)/k_B T_1)-1]^{-1}. \qquad (13)$$

Equation (12) describes an experimentally measurable intensity of thermal radiation of a moving particle and corresponds to the generalized Kirchhoff law: at total dynamical and thermal equilibrium ($\beta = 0$ and $T_1 = T_2$) we obtain $I = 0$, i. e. the balance between the absorption and emission. At $\beta \neq 0$, this balance is violated, and the particle emits more photons than absorbs. The emitted intensity of thermal radiation is given by Eq. (13), according to which the frequency-angular distribution takes the form (one has to put $x \equiv \cos\theta$ in (13) where $\theta$ is the angle between the emission direction and particle velocity)

$$\frac{d^2 I_1}{d\Omega\, d\omega} = \frac{\hbar\gamma\omega^4}{\pi^2 c^3} \frac{(1-\beta\cos\theta)^2 \alpha''(\omega\gamma(1-\beta\cos\theta))}{\exp(\hbar\omega\gamma(1-\beta\cos\theta)/k_B T_1)-1}. \qquad (14)$$

In what follows we discuss the case of the particle polarization corresponding to the low-frequency limit of the Drude dielectric permittivity: $\varepsilon(\omega) = i \cdot 4\pi\sigma_0/\omega$, where $\sigma_0$ is the static conductivity. Respectively, the electric and magnetic polarizabilities of a small particle of radius $R$ are given by [13]

$$\alpha_e''(\omega) = 3R^3\omega/4\pi\sigma_0, \qquad (15a)$$

$$\alpha_m''(\omega) = -\frac{3Rc^2}{8\pi\sigma_0\omega}\chi(x), \ x \equiv 2R\cdot(2\pi\sigma_0\omega)^{1/2}/c, \tag{15b}$$

$$\chi(x) = 1 - \frac{x}{2}\frac{\sinh x + \sin x}{\cosh x - \cos x}. \tag{16}$$

## 3. Contribution of the electric polarizability

We first consider the contribution of the electric polarizability $\alpha_e''(\omega)$. Using (15a), integrals (10)—(13) are calculated explicitly

$$F_x = -\frac{8\pi^4}{21}\frac{\hbar R^3}{c^4\sigma_0}\beta\left[\frac{(1+\beta^2/5)}{(1-\beta^2)}\vartheta_2^6 + \vartheta_1^6\right], \tag{17}$$

$$\dot{Q} = \frac{8\pi^4}{21}\frac{\hbar R^3}{c^3\sigma_0}\left[\frac{(1+2\beta^2+\beta^4/5)}{(1-\beta^2)}\vartheta_2^6 - (1-\beta^2)\vartheta_1^6\right], \tag{18}$$

$$I = \frac{8\pi^4}{21}\frac{\hbar R^3}{c^3\sigma_0}\left[\vartheta_1^6 - \frac{(1+\beta^2)}{(1-\beta^2)}\vartheta_2^6\right], \tag{19}$$

$$I_1 = \frac{8\pi^4}{21}\frac{\hbar R^3}{c^3\sigma_0}\vartheta_1^6, \tag{20}$$

where we have used the notation $\vartheta_i = k_B T_i/\hbar \ (i=1,2)$.

Integrating (14) over frequencies yields

$$\frac{dI_1}{d\Omega} = \frac{2\pi^3}{21}\frac{\hbar R^3 \vartheta_1^6}{c^3\sigma_0}\frac{(1-\beta^2)^2}{(1-\beta\cos\theta)^3}. \tag{21}$$

The angular intensity in the forward direction is given by

$$\left(\frac{dI_1}{d\Omega}\right)_{\theta=0} = \frac{2\pi^3}{21}\frac{\hbar R^3\vartheta_1^6\gamma^2}{c^3\sigma_0}(1+\beta)^3. \tag{22}$$

Since $dQ'/dt'$ represents the heating rate in the rest frame of the particle, with allowance for the relation $dt' = dt\cdot(1-\beta^2)^{1/2}$ we obtain

$$dQ'/dt' = C_0 dT_1/dt' = C_0(1-\beta^2)^{-1/2}dT_1/dt, \tag{23}$$

where $C_0$ and $T_1$ are the heat capacity and the particle temperature in $\Sigma'$. Using (5), (7) and (23) yields

$$dT_1/dt = \frac{\dot{Q}}{C_0(1-\beta^2)^{1/2}} = \frac{8\pi^4}{21} \frac{\hbar R^3}{c^3 \sigma_0 C_0} \left[ \frac{(1+2\beta^2+\beta^4/5)}{(1-\beta^2)^{3/2}} \vartheta_2^6 - (1-\beta^2)^{1/2} \vartheta_1^6 \right]. \qquad (24)$$

Since $\beta$ depends on time, Eq. (24) should be solved together with the dynamics equation

$$mc\,d\beta/dt = (1-\beta^2)^{3/2} F_x. \qquad (25)$$

In particular, from (25) and (17) it follows that (introducing the particle mass $m = \frac{4\pi}{3}\rho R^3$ with $\rho$ being the mass density)

$$d\beta/dt = -\frac{2\pi^3}{7} \frac{\hbar \vartheta_2^6}{c^5 \sigma_0 \rho} \beta \left[ \frac{1+\beta^5/5}{(1-\beta^2)} + \left(\frac{T_1}{T_2}\right)^6 \right] (1-\beta^2)^{3/2}. \qquad (26)$$

According to (26) and assuming $\sigma_0 = const$, the characteristic time scale of particle deceleration is given by

$$\tau_V = \frac{7}{2\pi^3} \frac{c^5 \sigma_0 \rho}{\hbar \cdot \vartheta_2^6}.$$

At the same time, Eq. (24) reduces to (introducing $C_0 = mC_s = \frac{4\pi}{3} R^3 \rho C_s$ with $C_s$ being the specific heat capacity)

$$\frac{d(T_1/T_2)}{dt} = \frac{2\pi^3}{7} \frac{k_B \vartheta_2^5}{c^3 \sigma_0 C_s \rho} \left[ \frac{(1+2\beta^2+\beta^4/5)}{(1-\beta^2)^{3/2}} - (1-\beta^2)^{1/2} \left(\frac{T_1}{T_2}\right)^6 \right] \qquad (27)$$

According to (27) and assuming $\sigma_0 = const$, $T_1 \neq T_2$, $T_2 = const$, the characteristic time scale of thermal relaxation is

$$\tau_Q = \frac{7}{2\pi^3} \frac{c^3 \sigma_0 C_s \rho}{k_B \cdot \vartheta_2^5},$$

and one obtains $\tau_Q / \tau_V = C_s T_2 / c^2 \ll 1$. Thus, for gold ($C_s = 129 J/kg \cdot K$, $T_2 = 300 K$) one obtains $\tau_Q \approx 10^3 s$, $\tau_Q / \tau_V = 4.3 \cdot 10^{-13}$ and the ratio $\tau_Q / \tau_V$ becomes still lower with decreasing $T_2$.

Such a large difference between $\tau_Q$ and $\tau_V$ implies that thermal state of particle and its temperature $T_1$ can be calculated assuming that $\beta \cong const$. This is confirmed by more accurate numerical calculations in Sect. 4.

In particular, from (27) (if $\sigma_0$ is temperature independent) it follows that temperature $T_1$ of the particle reaches the stationary value depending on $T_2$ and meets the steady-state conditions $\dot{Q} = 0$ and $dT_1 / dt = 0$:

$$T_1 = T_2 \left( \frac{1 + 2\beta^2 + \beta^4/5}{(1-\beta^2)^2} \right)^{1/6} \tag{28}$$

This mechanism is restricted by the condition $T_1 < T_m$, where $T_m$ is the melting point of the particle. Then at $\gamma \gg 1$ we obtain $\gamma \leq (T_m / 1.2 T_2)^{3/2}$. The process of particle evaporation and ionization (at higher temperatures) needs a special consideration.

Substituting (28) in (17)--(20) yields

$$F_x = -\frac{16\pi^4}{21} \frac{\hbar R^3 \vartheta_2^6}{c^4 \sigma_0} \beta \frac{(1 + 0.6\beta^2)}{(1-\beta^2)^2} \tag{29}$$

$$I = \frac{16\pi^4}{21} \frac{\hbar R^3 \vartheta_2^6}{c^3 \sigma_0} \beta^2 \frac{(1 + 0.6\beta^2)}{(1-\beta^2)^2} \tag{30}$$

$$I_1 = \frac{8\pi^4}{21} \frac{\hbar R^3 \vartheta_2^6}{c^3 \sigma_0} \frac{(1 + 2\beta^2 + \beta^4/5)}{(1-\beta^2)^2} \tag{31}$$

Moreover, the steady-state angular intensity of radiation is given by

$$\frac{dI_1}{d\Omega} = \frac{2\pi^3}{21} \frac{\hbar R^3 \vartheta_2^6}{c^3 \sigma_0} \frac{(1 + 2\beta^2 + \beta^4/5)}{(1 - \beta\cos\theta)^3} \qquad (32)$$

$$\left(\frac{dI_1}{d\Omega}\right)_{\theta=0} = \frac{2\pi^3}{21} \frac{\hbar R^3 \vartheta_2^6 (1 + 2\beta^2 + \beta^4/5)(1+\beta)^3 \gamma^6}{c^3 \sigma_0} \qquad (33)$$

During steady-state motion of the particle kinetic energy is entirely converted into radiation. Therefore, the total energy emitted up to stopping is equal to $(\gamma_s - 1)mc^2$, where $\gamma_s$ corresponds to the instant of the onset of equilibrium.

## 4. Numerical calculations

Studying the contribution of the magnetic polarizability needs a numerical calculation, since the nanosized particles made of good conductors have large magnetic polarization [8,14]. For a numerical example we have chosen the case of gold particle using Eqs. (10)—(13) with allowance for electric $\alpha_e(\omega)$ and magnetic $\alpha_m(\omega)$ polarizabilities (15) and (16). Since the conductivity $\sigma_0$ and specific heat capacity $C_s$ of gold are strongly dependent on temperature $T_1$, the tabulated dependences of these parameters were taken from [15].

The equilibrium ratio $T_1/T_2$ was numerically calculated from Eq. (11) under the condition $\dot{Q} = 0$ and the resultant dependences of $T_1/T_2$ on the Lorentz-factor $\gamma$ are shown in Fig. 2. It should be emphasized once again, that the calculated steady-state temperature $T_1$ corresponds to the own reference frame of the particle (co-moving reference frame). As one can see from Fig. 2, the ratio $T_1/T_2$ decreases with increasing the particle radius. This is the consequence of specific dependence of the magnetic polarizability on the particle radius. The corresponding dependence on temperature $T_2$ is not so noticeable at a fixed $R$. Moreover, we can conclude that at $\gamma \gg 1$ the dependence of $T_1/T_2$ is described by Eq. (28) (solid line) for particles with a radius of several nanometers.

Assuming $\beta = const$ and $T_2 = const$, we calculated the dependence $T_1(t)$ and the time $\tau$ needed to reach the equilibrium temperature $T_1^{(eq)}$, provided that the initial particle temperature $T_1$ is known. In this case the equation similar to Eq. (27) with allowance for the magnetic polarization contribution in the right-hand side was solved numerically. Figure 3a shows the calculated values of $\tau$ depending on the Lorentz factor $\gamma$.

. In Fig. 3a, the bottom curve (symbols "+") is interrupted since the particle is heated and its temperature reaches the melting point with further increase in $\gamma$. As we can see, the value of $\tau$ depends on the radiation temperature of vacuum and the particle radius and varies in a rather wide interval from $10^{-4}$ to $10^8 \, s$. Shorter times $\tau$ for bigger particles (circles and crosses) are explained by the lower values of $T_1/T_2$ at equilibrium (Fig. 2).

Figure 3b shows the time of fading ($\tau_{1/e}$) the particle velocity by $e$ times depending on the initial relativistic factor $\gamma_0$ and assuming different thermal conditions. The solid and dashed curves correspond to the conditions of constant temperature of the particle ($300K$) and a vacuum background ($2.7K$), and a particle radius of $5nm$ (solid line) and $50nm$ (dashed line). The dashed-dotted and dotted lines correspond to the case where the particle temperature reaches a steady-state value (Fig. 2) corresponding to a vacuum radiation temperature of $2.7K$ (dashed-dotted line) and $300K$ (dotted line), and the particle radius of $5nm$. The dotted line is interrupted since the particle temperature reaches the melting point. Comparing the times $\tau$ and $\tau_{1/e}$ confirms the conclusion in Sect. 3 that the process of thermal relaxation proceeds much faster than the process of deceleration (cf. the interrupted lines in Figs. 3a and 3b, the solid line in Fig. 3b and the line shown by squares in Fig. 3a). It is worth noting that the upper two lines in Fig. 3b demonstrate the minimum values of $\tau_{1/e}$ for the chosen initial conditions while the thermal relaxation significantly increases the time of velocity fading.

Figure 4a,b shows the calculated intensities of the steady-state thermal radiation (a) and absorption (b) normalized to $R^3$ ($I_1/R^3$ and $I_2/R^3$, respectively). Unlike temperature $T_1$ corresponding to the reference frame $\Sigma'$, the aforementioned intensities correspond to the reference frame of vacuum, $\Sigma$. The lines shown by symbols "+" correspond to only the contributions from the electric polarization (Eqs. (30), (31)). Due to the $R^{-3}$ normalization, the latter curves are universal and do not depend on the particle radius. From Fig. 4 we can conclude that in the case of good metals (such as gold), the contribution of the magnetic polarization dominates in the thermal radiation intensity of moving particle and it should necessarily be taken into consideration for particles with radii from units to tens nanometers.

Figure 5 shows the ratio $I_1/I_2$ of the steady-state emission and absorption intensities (in the reference frame of vacuum). Solid line is the fitting dependence $2\gamma^{1.98}$, which is very close to the dependence $I_1/I_2 = \gamma^2$ following from Eqs. (30) and (31) at $\gamma \gg 1$ (i.e. with allowance for only the electric polarization). In the limit $\beta \to 0$, the difference between $I_1$ and $I_2$ tends to zero and does not depend on the particle radius as it should be at equilibrium. For example, at $\beta = 0.1$ one obtains $(I_1 - I_2)/I_1 = 0.015$.

## 4. Summary and conclusions

To conclude, we have obtained a complete set of equations describing the fluctuation-electromagnetic interaction of a small polarizable particle moving with an arbitrary velocity through the vacuum background of a certain temperature: characteristics of the vacuum-assisted radiation, friction force and the rate of heating (cooling) of the particle. We have shown that the particle in the co-moving frame acquires the radiation temperature of vacuum, multiplied by the function of the Lorentz-factor $\gamma$, and then emits thermal photons predominantly in the forward direction up to stopping. The time of thermal relaxation is considerably smaller than the time of stopping and depends on the initial conditions. The intensity of emission (at steady-state conditions with $\gamma \gg 1$) is much higher than for a particle at rest. Thus, the ratio of the emitted and absorbed radiation power is proportional to $\sim \gamma^2$. For metallic particles with high conductivity the contribution from the magnetic polarizability dominates. Our results do not contradict the idea that there is no universal relativistic temperature transformation [2].

In addition to the basic theoretical value, the results may be of great interest for the laboratory experiments related to creating new sources of directional microwave radiation, particle trapping in cavities, and for astrophysics. Astrophysical applications can be associated with studying the evolution of comet tails, observation of microwave cosmic radiation upon gravitational compression of gas and dust clouds and accretion onto massive cosmic objects. Directional effect of thermal radiation of nanosized moving particles can affect an observed anisotropy of the primary 2.7 K blackbody radiation.

**Acknowledgments.** We thank Prof. G. Barton and Dr. V. E. Mkrtchian for fruitful comments and constructive criticism.

## References

[1] G. A. Kluitenberg, S. R. De Groot, and P. Mazur, Physica 19 (1553) 689; ibid. 19 (1953) 1079; G. A. Kluitenberg, Relativistic Thermodynamics of Irreversible Processes (Proefschrift, Leiden, 1953).

[2] E. Bormashenko, Entropy 9 (2007) 113.

[3] G. Ares de Parga, B. Lopez-Carrera, and F. Angulo-Brown, J. Phys. A.: Math. Gen. 38 (2005) 2821; B. Lopez-Carrera, M. A. Rosales, and G. Ares de Parga, Chin. Phys., B 19 (4) (2010) 040203; Zhong Chao Wu, Europhys. Lett., 88 (2009) 20005; T. S. Biro and P. Van, Europhys. Lett. 89 (2010) 30001.

[4] G. W. Ford and R. F. O'Connell, Phys. Rev. E, 88 (2013) 044101.

[5] G. R. Henry, R. B. Feduniak, J. E. Silver, and M. A. Paterson, Phys. Rev. 176 (1968) 1451; V. R. Balasanyan and V. E. Mkrtchian, Armen. J. Phys. 2 (2009) 182.

[6] J. D. Jackson, Classical electrodynamics (Wiley, New York—London, 1962)

[7] G. V. Dedkov and A. A. Kyasov, Phys. Solid State, 45 (2003) 1815.

[8] G. V. Dedkov and A. A. Kyasov, J. Phys.: Condens. Matter, 20 (2008) 354006.

[9] G. V. Dedkov and A. A. Kyasov, Physics Letters A. 339 (2005) 212.

[10] G. V. Dedkov and A. A. Kyasov, Nucl. Instr. Meth. B268 (2010) 599.

[11] V. E. Mkrtchian, V. A. Parsegian, R. Podgornik, and W. M. Saslow, Phys. Rev. Lett. 91 (2003) 220801.

[12] F. Intravaia, C. Henkel, and M. Antezza, in: Casimir Physics (Lecture Notes in Physics, 834 (2011) 345), ed. D.A.R. Dalvit, P.W. Milonni, D. Roberts and F. Da Rosa (Berlin: Springer), Ch. 11, pp. 345-391.

[13] L. D. Landau, E. M. Lifshitz, Electrodynamics of Continuous Media (Pergamon, Oxford, 1960).

[14] P.-O. Chapuis, M. Laroche, S. Volz, and J.-J. Greffet, Appl. Phys. Lett. 92 (2008) 201906; A. Manjavacas and F. J. Garcia de Abajo, Phys. Rev. B86 (2012) 075466.

[15] Physical Quantities. Handbook, ed. by I. C. Grigoriev and E. Z. Meilikhov (Atomizdat, Moscow, 1991) (in Russian).

## FIGURE CAPTIONS

Fig. 1. The reference systems of vacuum ($\Sigma$) and particle ($\Sigma'$). $\mathbf{S_1}$ and $\mathbf{S_2}$ denote the Poynting vectors of the emitted and absorbed radiation, $\sigma$ is the wave-surface.

Fig. 2 Equilibrium ratio $T_1/T_2$ depending on $\gamma$. Squares: $R=5nm, T_2=2.7K$; circles: $R=50nm, T_2=2.7K$; rhombs: $R=50nm, T_2=50K$; crosses: $R=5nm, T_2=50K$. Numerical calculation was performed with allowance for the electric and magnetic particle polarization. Solid line: calculated by Eq. (28).

Fig. 3a. Time establishing a steady thermal state depending on the Lorentz-factor $\gamma$ at different initial conditions. Squares: $T_1=300K, T_2=2.7K, R=5nm$; circles: $T_1=300K, T_2=2.7K, R=50nm$; rhombs: $T_1=2.7K, T_2=50K, R=5nm$; crosses: $T_1=2.7K, T_2=50K, R=50nm$; symbols +: $T_1=2.7K, T_2=300K, R=5nm$. The interrupted line corresponds to the case where the particle reaches the point of melting.

Fig. 3b. Time of fading ($\tau_{1/e}$) the particle velocity by $e$ times depending on the initial relativistic factor $\gamma_0$ and different thermal conditions. Solid line: $T_1=300K; T_2=2.7K; R=5nm$; dashed line: $T_1=300K; T_2=2.7K; R=50nm$; dashed-dotted line: $T_1=T_1^{(eq)}; T_2=2.7K; R=5nm$; dotted line: $T_1=T_1^{(eq)}; T_2=300K; R=5nm$.

Fig. 4a,b. Dependences of the reduced steady-state emission intensity $I_1/R^3$ (a) and absorption intensity $I_2/R^3$ (b) on the Lorentz factor $\gamma$. Squares: $T_2=2.7K, R=5nm$; circles: $T_2=2.7K, R=50nm$; rhombs: $T_2=50K, R=5nm$; crosses: $T_2=50K, R=50nm$; symbols +: contribution from the electric particle polarization only ($T_2=2.7K, R=50nm$).

Fig. 5. The ratio of the intensity of radiation to the intensity of absorption at steady-state conditions. Squares: $T_2=2.7K, R=5nm$; crosses: $T_2=2.7K, R=50nm$; symbols +: $T_2=50K, R=5nm$; solid line –fitting curve $I_1/I_2=2\gamma^{1.95}$.

FIGURE 1

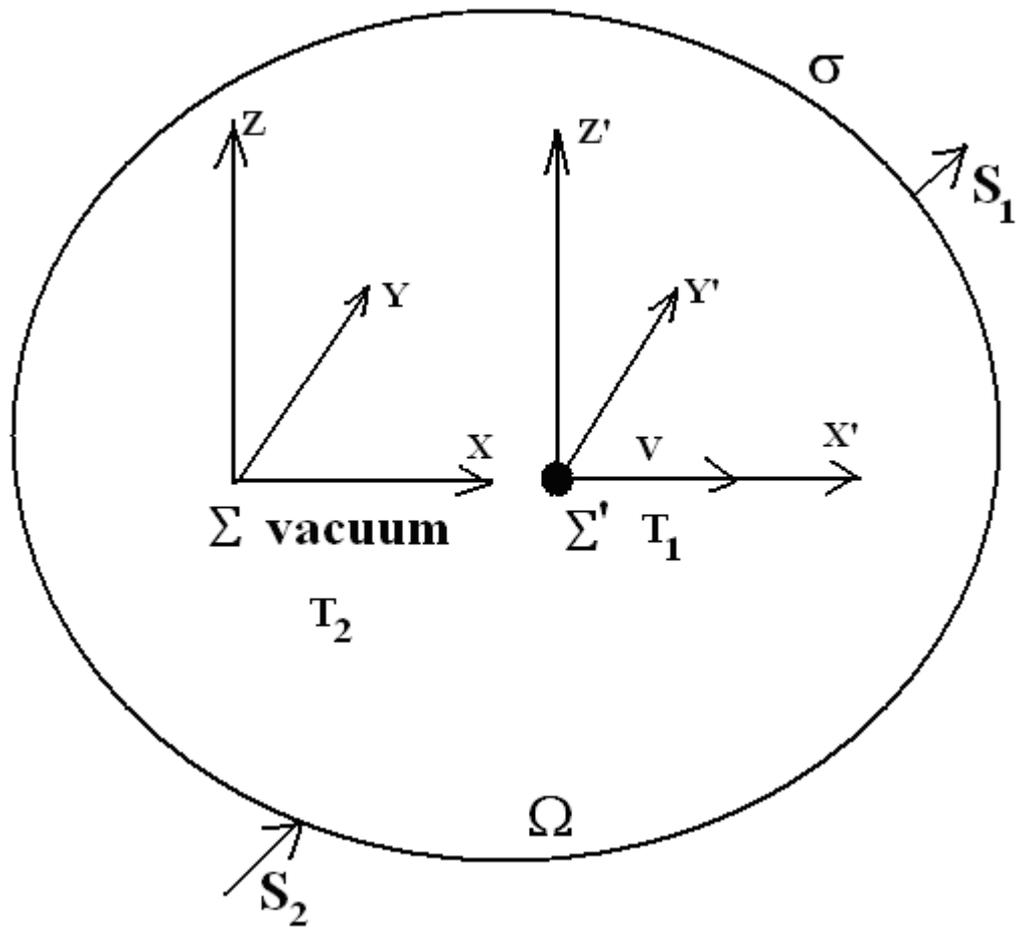

FIGURE 2

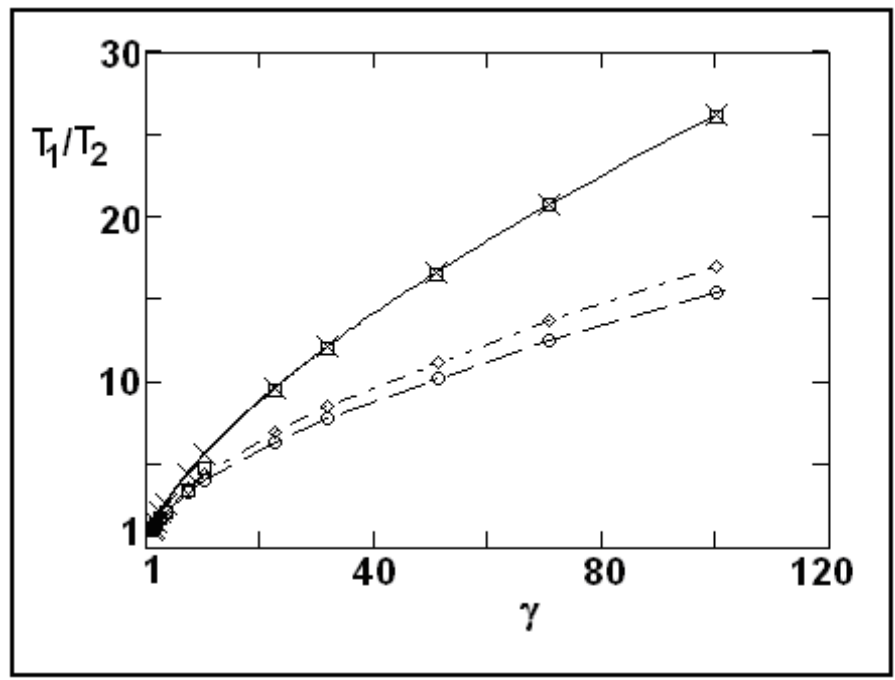

15FIGURE 3a

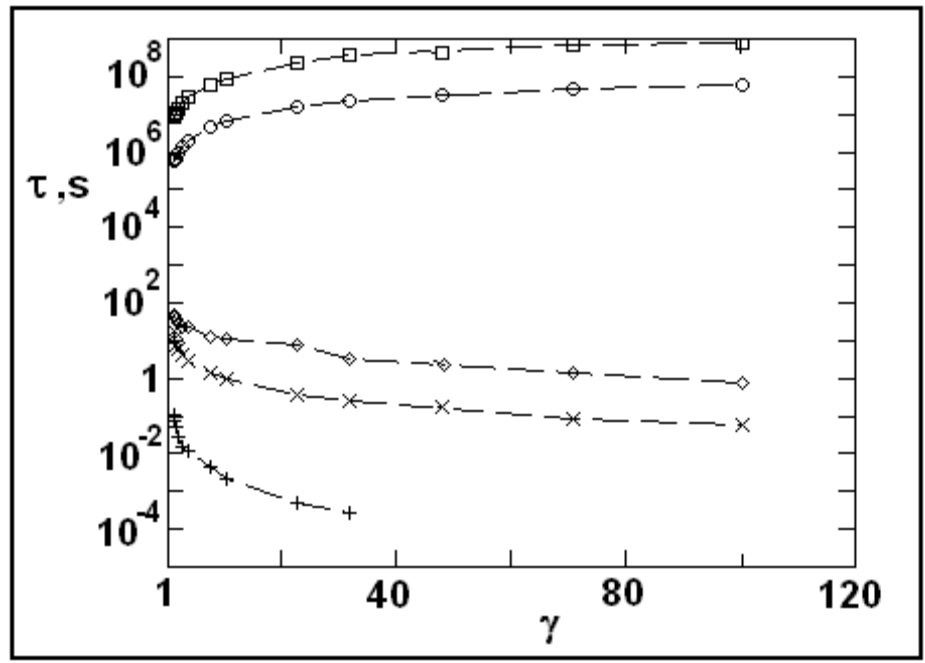

FIGURE 3b

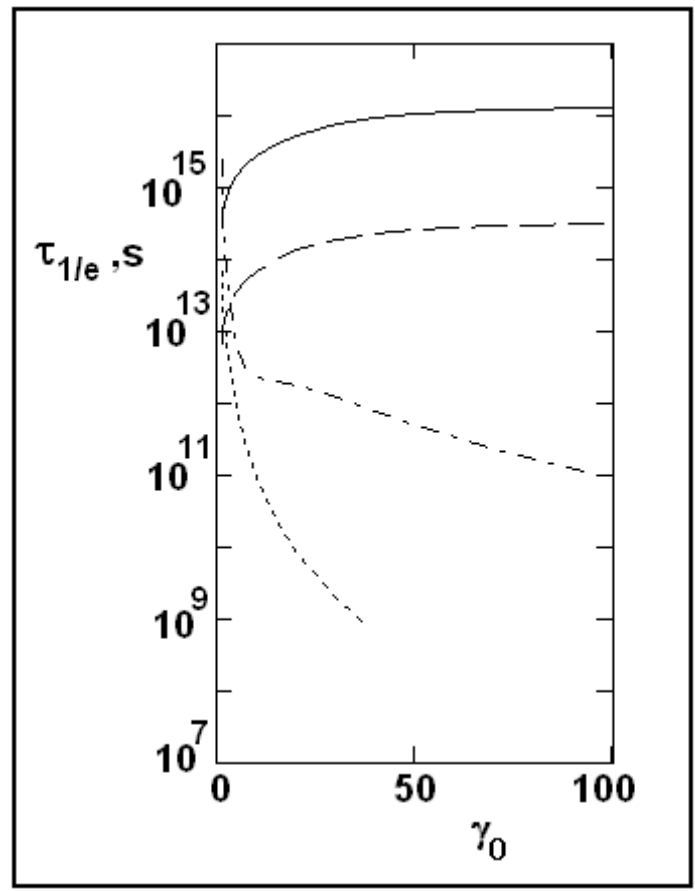

FIGURE 4a

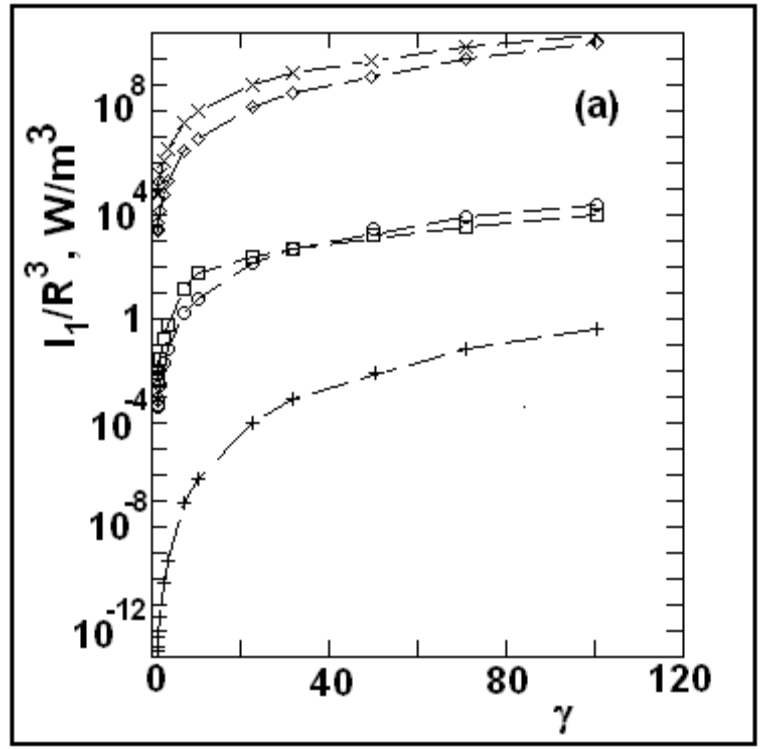

FIGURE 4b

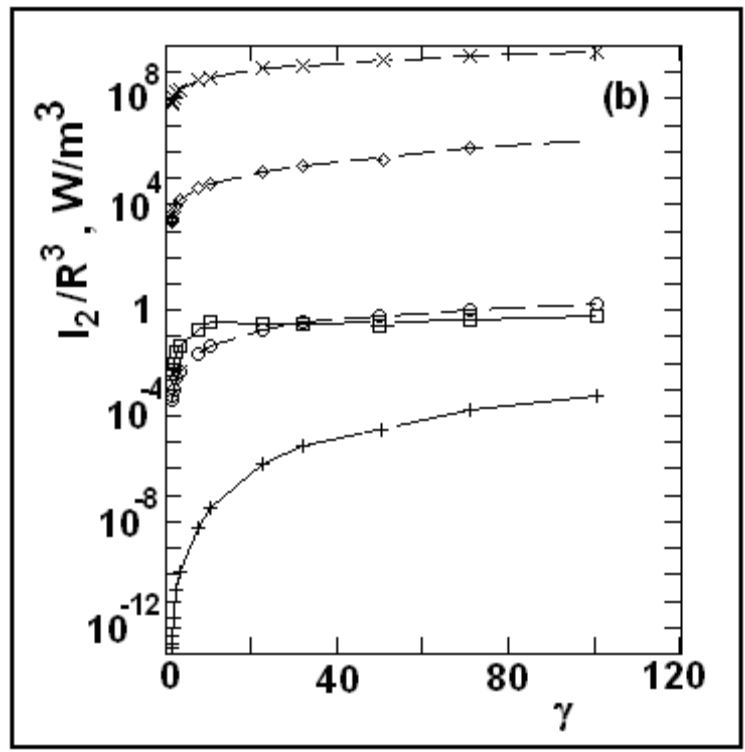

FIGURE 5

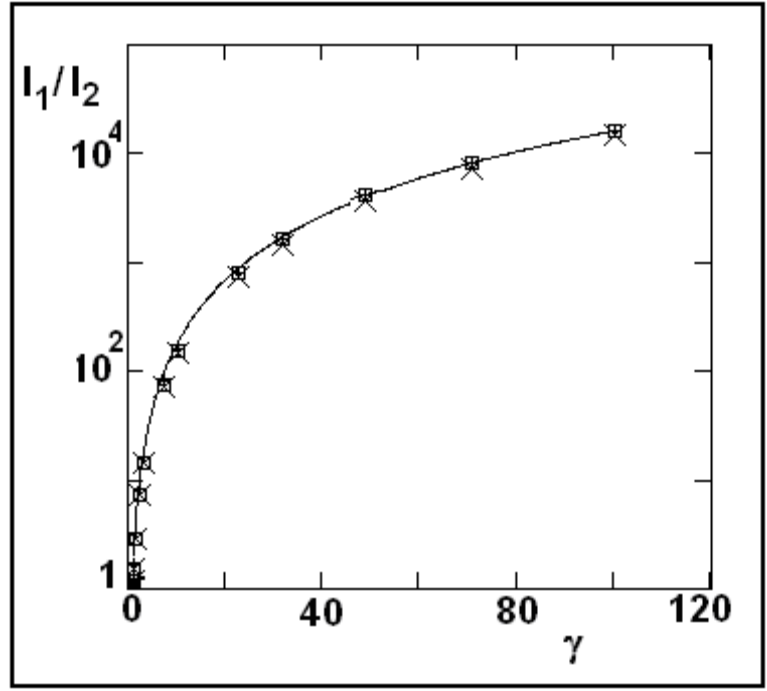